\documentclass[aps,fleqn,amsmath,amssymb]{revtex4}

\newcommand{\bed}{\[}
\newcommand{\eed}{\]}
\newcommand{\beq}{\begin{equation}}
\newcommand{\eeq}{\end{equation}}
\newcommand{\beqa}{\begin{eqnarray}}
\newcommand{\eeqa}{\end{eqnarray}}
\newcommand{\ket} [1] {\vert #1 \rangle}
\newcommand{\bra} [1] {\langle #1 \vert}
\newcommand{\braket}[2]{\langle #1 | #2 \rangle}
\newcommand{\proj}[1]{\vert{#1} \rangle \langle {#1} \vert}

\newcommand{\gras}[1]{\mathbf{#1}}

\usepackage{graphicx}
\usepackage[mathscr]{eucal}
\usepackage{amsmath}
\usepackage{amssymb}
\usepackage{epstopdf}
\usepackage{epsfig}
\begin{document}

\title{Optimal finite measurements and Gauss quadratures}
\author{Sofyan Iblisdir$^1$}
\author{J\'er\'emie Roland$^{2,3}$}
\affiliation{$^1$ GAP-Optique, Universit\'e de Gen\`eve, 1211 Gen\`eve, Switzerland \\
$^2$ Quantum Information and Communication, Ecole Polytechnique, CP 165/59, Universit\'e Libre de Bruxelles, 1050 Bruxelles, Belgium\\
$^3$ Laboratoire de Recherche en Informatique, UMR 8263, Universit\'e de Paris-Sud, 91405 Orsay, France}
\email{sofyan.iblisdir@physics.unige.ch, jroland@ulb.ac.be}

\date{\today}

\begin{abstract}
We exhibit measurements for optimal state estimation which have a finite number of outcomes. This is achieved by a connection between finite optimal measurements and Gauss quadratures. The example we consider to illustrate this connection is that of state estimation on $N$ qubits, all in a same pure state. Extensions to state estimation of mixed states are also discussed.
\end{abstract}

\maketitle

Two central questions in quantum information theory are: "How much information can be encoded in a finite quantum ensemble?" and "What measurements should one use to optimally retrieve this information?". There are many ways to address these questions. One of the simplest is that proposed in \cite{mass95}, in the form of a game between a referee, Alice, and a player, Bob. Alice prepares $N$ qubits in an identical state $\ket{\Omega}$, unknown to Bob. This state is drawn from a uniform probability distribution on the set of all pure qubit states, the so-called Bloch sphere. Then Bob performs some measurement on $\ket{\Omega}^{\otimes N}$, to guess the prepared state $\ket{\Omega}$. If the state guessed by Bob is  $\ket{\Omega_k}$, he achieves the score $|\braket{\Omega}{\Omega_k}|^2$. Bob aims at achieving the best possible average score
\beq
\Sigma_N=\sum_k \int d\Omega \; P_N(\Omega_k|\Omega) \;  |\braket{\Omega}{\Omega_k}|^2,
\eeq
where $\int d\Omega=1$ and where $P_N(\Omega_k|\Omega) $ denotes the conditional probability of Bob to get a measurement outcome  $\ket{\Omega_k}$ when Alice has prepared the state $\ket{\Omega}^{\otimes N}$.

It is proven in \cite{mass95}, that when $N$ identical copies of the qubits whose state we want to estimate are available, the optimal value that $\Sigma_N$ can take is 
\beq\label{eq:optscore}
\Sigma^{\textrm{opt}}(N)=\frac{N+1}{N+2}.
\eeq
Also, one readily checks that the following positive-operator-valued measure (povm) achieves the optimal score:
\beq
(N+1) \ket{\Omega}\bra{\Omega}^{\otimes N} d\Omega.
\eeq
In the sequel, we will call this povm the \emph{coherent} povm, because it is made of projectors onto the so-called SU(2) coherent states \cite{perelomov}. While this povm requires an infinite number of outcomes, it is possible to build povm's achieving the optimal score $\Sigma_N^{\textrm{opt}}$ and whose number of outcomes is \emph{finite}. This problem has been addressed by several authors. In \cite{derk98}, an algorithm to derive finite povm's achieving the score (\ref{eq:optscore}) was proposed. Later, in \cite{lato98}, the issue of finding minimal optimal finite povm's was tackled and a solution was provided for up to 7 qubits. 

The present note aims at bringing a new perspective on this problem. We show that there is a natural connection between this problem and the design of a Gauss quadrature on a sphere (see also \cite{baga01}). Let us first introduce some notations and reformulate the game between Alice and Bob in more precise terms. In the following, $\{\ket{0},\ket{1\}}$ will denote an orthonormal basis of the Hilbert space of a qubit. Pure qubit states will be parametrised as 
\beq
\ket{\Omega}=\cos{\frac{\theta}{2}} \ket{0}+ e^{i \phi} \sin{\frac{\theta}{2}} \ket{1}.
\eeq
where $0 \leq \phi \leq 2\pi$ and $0 \leq \theta \leq \pi$. $\Omega=(\theta,\phi)$ denotes the direction on the Bloch sphere associated with the qubit state. 

The smallest Hilbert space describing the states $\ket{\Omega}^{\otimes N}$ is the symmetric subspace
of the Hilbert space of $N$ qubits, $\mathscr{H}_+^{\otimes N}$ (that is the subspace of maximal total spin $N/2$).
As any measurement on $\mathscr{H}_+^{\otimes N}$, the measurement used by Bob is a povm describable by a set of operators
$\{\mathscr{O}_k\}$ satisfying 
\beq\label{eq:condpovm-1}
\mathscr{O}_k \geq 0, 
\eeq
\beq\label{eq:condpovm-2}
\sum_k \mathscr{O}_k=\gras{I}_N,
\eeq
where $\gras{I}_N$ denotes the identity on $\mathscr{H}_+^{\otimes N}$.

Now observe that any set of $n$ operators $\mathscr{O}_k=c_k \ket{\Omega_k}\bra{\Omega_k}^{\otimes N}$ such that 
\beq\label{eq:condpovm}
\sum_{k=1}^{n} c_k \ket{\Omega_k}\bra{\Omega_k}^{\otimes N}=\gras{I}_N.
\eeq
is an optimal povm. Indeed this povm is such that $P_N(\Omega_k | \Omega)=c_k|\braket{\Omega_k}{\Omega}|^{2N}$, it achieves the score
\bed
\Sigma_N(\{\mathscr{O}_k\})=\sum_{k=1}^{n} c_k \int d\Omega |\braket{\Omega_k}{\Omega}|^{2(N+1)}.
\eed
As a consequence of Shur's lemma, we have
\bed
\int d\Omega \ket{\Omega}\bra{\Omega}^{\otimes N+1}=\frac{\gras{I}_{N+1}}{N+2}.
\eed
Thus the score achieved by the povm (\ref{eq:condpovm}) reads
\beq
\Sigma_N(\{\mathscr{O}_k\})=\frac{1}{N+2} \sum_{k=1}^{n}  c_k =\frac{N+1}{N+2}=\Sigma_N^{\textrm{opt}}.
\eeq
The issue of finding optimal finite povm's can actually be re-stated in a very simple form, that allows for a straightforward derivation with $n=(N+1) \lceil(N+1)/2 \rceil$ for arbitrary values of $N$, as well as other more "economical" optimal povm's for some large $N$ values. Our construction is based on the concept of Gauss quadrature which we briefly describe now.

A Gauss quadrature is a rule designed to approach an integral over some domain $\mathscr{D}$ by a sum and defined by a mesh of points $x_k \in \mathscr{D}$ and associated weights $\lambda_k$,
\beq
\int_{\mathscr{D}} dx f(x) \approx \sum_k \lambda_k f(x_k).
\eeq
If the parameters $\{x_k,\lambda_k\}$ are chosen so that the quadrature is exact for a set of functions $f_j(x)$:
\beq
\int_{\mathscr{D}} dx f_j(x)= \sum_k \lambda_k f_j(x_k),
\eeq
it will of course be exact for any linear combination of the form $f(x)=\sum_j \alpha_j f_j(x)$. 

In order to associate a Gauss quadrature with our problem, let us observe that Eq.(\ref{eq:condpovm}) can be expressed as
\beq\label{eq:condpovm2}
\sum_{k=1}^n c_k |\braket{\Omega_k}{\Omega}|^2=1 \hspace{1cm} \forall \Omega.
\eeq

Expanding $\braket{\Omega_k}{\Omega}$ in terms of spherical harmonics $Y_l^m(\Omega)$, this condition can be re-expressed as the following system of equations \cite{lato98}:
\begin{subequations}\label{eq:system}
\beqa
\sum_{k=1}^{n}c_{k}&=&N+1\\
\sum_{k=1}^{n}c_{k} Y_l^m(\Omega_k)&=&0,
\eeqa
\end{subequations}
where $l=1,\dots,N; m=-l,\dots,l$. Using the following properties of spherical harmonics:
\beq
Y_0^0(\Omega)=1/\sqrt{4\pi},
\eeq
\beq
\int d\Omega Y_l^{m*}(\Omega) Y_{l'}^{m'}(\Omega)=\delta_{ll'} \delta_{m m'},
\eeq
we see that the conditions (\ref{eq:condpovm2}) can be rewritten as
\beq\label{eq:main}
\int d\Omega Y_l^m(\Omega)=\sum_{k=1}^n \lambda_k Y_l^m(\Omega_k),
\eeq
where $\lambda_k=4\pi c_k/(N+1)$. Thus Eq.(\ref{eq:main}) means that finding an optimal finite povm amounts to finding an exact Gauss quadrature on the sphere for the spherical harmonics up to order $l=N$.

We now show that some quadratures can be found very simply. Spherical harmonics can be written as:
\beq
Y_l^m(\Omega)= C_l^m P_l^{|m|}(\cos\theta)e^{im\phi}
\eeq
where $C_l^m$ are constants (whose precise expression are irrelevant here),
\beq
P_l^m(u)=(1-u^2)^{\frac{m}{2}}\frac{d^m}{du^m}P_l(u),
\eeq
are the associated Legendre functions and
\beq
P_l(u)=\frac{1}{2^ll!}\frac{d^l}{du^l}(u^2-1)^l
\eeq
are the Legendre polynomials. We see that the spherical harmonics are either a product of $e^{im\phi}$ and a polynomial of degree $l$ in $\cos\theta$ (when $m$ is even), either a product of $e^{im\phi}$, $\sin\theta$ and a polynomial of degree $l-1$ in $\cos\theta$ (when $m$ is odd).

Thus, expressing Eq.(\ref{eq:main}) in spherical coordinates, we must find a Gauss quadrature that is exact for:
\beq
\int_{-1}^{1} d(\cos\theta)  \int_0^{2 \pi} d\phi \cos^n\theta e^{im\phi}
\eeq
$\forall \ m$ even ($0\leq m \leq N$) and $n$ ($0\leq n \leq N$), and exact for
\beq
\int_{-1}^{1} d\cos\theta  \int_0^{2 \pi} d\phi \sin\theta \cos^n\theta e^{im\phi}
\eeq
$\forall \ m$ odd ($0\leq m \leq N$) and $n$ ($0\leq n \leq N-1$).

A simple Gauss quadrature is obtained by separating integrations over the variables $\theta$ and $\phi$
and defining an independent Gauss quadrature for each. Let us first consider the integral in $\phi$. Since $m$ is integer,
we have:
\beq\label{mneq0}
\int_0^{2\pi} d\phi\;  e^{im\phi}= \delta_{m0}
\eeq
Since $m$ can take integer values up to $N$, it is easy to see that the minimal Gauss quadrature that will be exact for these $N+1$ integrals will require $n_1=N+1$ equidistant points, for instance the $(N+1)^{\text{th}}$ roots of unity $\phi_j=k\frac{2\pi}{N+1} \ (0\leq j \leq N)$, each with a same weight $\frac{2\pi}{N+1}$.

Let us now consider the integral in $\theta$. Eq.(\ref{mneq0}) shows that integrals for $m\neq 0$ will cancel out if the quadrature for $\phi$ is exact. So, the quadrature for $\theta$ needs only be exact for $m=0$, that is for the integrals 
\beq
\int_{-1}^1 d(\cos\theta) \cos^n\theta=\int_{-1}^1 dx x^n \ \forall \ 0\leq n \leq N.
\eeq
The following theorem \cite{szego} provides us with such quadratures: 

\newtheorem{gaussquad}{Theorem} 
\begin{gaussquad}\label{thm:gaussquad}
Let $[a,b] \subset \gras{R}$, and let $\{p_n\}$ denote a complete set of orthogonal polynomials on $\gras{L}^2([a,b])$. If $x_1 < \ldots < x_n$ denote the zeros of $\{p_n(x)\}$, there exist real numbers $\lambda_1, \ldots, \lambda_n$ such that 
\beq
\int_a^b d\alpha(x) \rho(x)=\sum_{k=1}^{n} \lambda_k \rho(x_k),
\eeq
whenever $\rho(x)$ is an arbitrary polynomial of degree at most $2n-1$. Moreover, the distribution $d\alpha(x)$ and the integer $n$ uniquely determine these numbers $\lambda_k$.
\end{gaussquad}
Using this theorem allows straightforwardly to build a Gauss quadrature on $[-1,1]$ using $n_2=\lceil (N+1)/2\rceil$ points, that is exact for any polynomial up to degree $N$,
such that we finally obtain a povm with
\beq
n=n_1 n_2=(N+1)\left\lceil \frac{N+1}{2}\right\rceil
\eeq
elements. As the $\theta$-part of the quadrature is based on Legendre polynomials, this separated variables Gauss quadrature is generally known as Gauss-Legendre quadrature.

The derivation of optimal povm's from the Gauss-Legendre quadrature is generic but we have no guarantee that it provides us with a \emph{minimal} optimal povm. Actually, it is not the case, as we already know that Latorre {\em et al} have been able to construct optimal finite povm's with less elements than ours, at least for $N\leq 7$ \cite{lato98}. One could have expected that the Gauss-Legendre quadrature we have presented above is not minimal. Heuristically, finding a minimal Gauss quadrature is equivalent to finding $n$ points on a sphere so as to optimally cover it. Intuitively, one expects that minimality  will in general not be achieved by considering the 2 variables $\theta$ and $\phi$ separately as we did. In a series of articles \cite{lebe92,lebe94,lebe99}, Lebedev deals with the issue of defining "good" quadratures and provided a quadrature rule on the sphere for all odd values up to $N=29$, as well as for higher values in the form $N=6a+5$ up to $N=131$ (or $a=21$). This last sequence requires a number of points
\beq
n=\frac{(N+1)^2}{3}+2,
\eeq
improving on Gauss-Legendre (separated variables) quadrature and on the conjectured minimal number of points given by Eq.~(24) of \cite{lato98}. 

\begin{figure}[h]
\begin{center}
\epsfig{figure=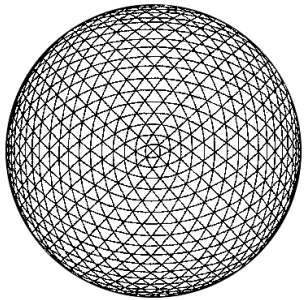,width=40mm,height=40mm}
\caption{Lebedev quadrature of order $N=59$. Distribution of $1800$ points for Lebedev quadrature of order $N=59$ (taken from \cite{lebe94}).}\label{lebedev}
\end{center}
\end{figure}

Fig.~\ref{lebedev} shows one of the quadratures proposed by Lebedev in \cite{lebe94}.
As may be seen, quadratures may be "polarised", in the sense that not all points are equivalent by symmetry,
moreover not all points are associated with a same weight.

\begin{figure}[h]
\begin{center}
\epsfig{figure=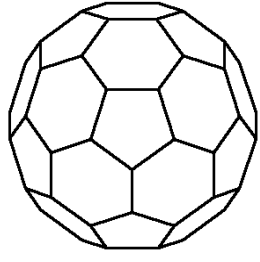,width=40mm,height=40mm}
\caption{Spherical design with $n=60$ points defining a Gauss quadrature on the sphere of order $N=10$ \cite{hard96}.}\label{spherical}
\end{center}
\end{figure}

A uniform distribution on the sphere is actually achievable for some number of points $n$, and is called a spherical $N$-design if the Gauss quadrature
it defines is exact for all spherical harmonics up to order $N$:
\beq
\int d\Omega \; Y_l^m(\Omega)=\frac{4\pi}{n}\sum_{k=1}^n Y_l^m(\Omega_k), l=1 \ldots N, m=-l \ldots l.
\eeq
Spherical $N$-designs are known up to $N=13$ \cite{hard96}. An example is displayed in Fig.~\ref{spherical}.  

A comparison between the various quadratures we have discussed is shown in the four first columns of Table~\ref{comparison}.

\begin{table}[p]
\begin{center}
\begin{tabular}{|c||c|c|c|c||c|c|}
\hline
$N$
& \rotatebox{90}{Latorre {\em et al}}
& \rotatebox{90}{Legendre}
& \rotatebox{90}{Lebedev}
& \rotatebox{90}{spherical design}
& \rotatebox{90}{Legendre (mixed)}
& \rotatebox{90}{Lebedev (mixed)}\\
\hline
$1$ & $2$ & $2$ &  & $2$ & $2$ & $2$ \\
$2$ & $4$ & $6$ &  & $4$ & $7$ & $ $ \\
$3$ & $6$ & $8$ & $6$ & $6$ & $10$ & $8$ \\
$4$ & $10$ & $15$ &  & $12$ & $22$ & $ $ \\
$5$ & $12$ & $18$ & $14$ & $12$ & $28$ & $22$ \\
$6$ & $18$ & $28$ &  & $24$ & $50$ & $ $ \\
$7$ & $22$ & $32$ & $26$ & $24$ & $60$ & $48$ \\
$8$ &  & $45$ &  & $36$ & $95$ & $ $ \\
$9$ &  & $50$ & $38$ & $48$ & $110$ & $86$ \\
$10$ &  & $66$ &  & $60$ & $161$ & $ $ \\
$11$ &  & $72$ & $50$ & $70$ & $182$ & $136$ \\
$12$ &  & $91$ &  & $84$ & $252$ & $ $ \\
$13$ &  & $98$ & $74$ & $94$ & $280$ & $210$ \\
$14$ &  & $120$ &  &  & $372$ & $ $ \\
$15$ &  & $128$ & $86$ &  & $408$ & $296$ \\
$\vdots$ & & $\vdots$ & $\vdots$ & & $\vdots$ & \\
\hline
\end{tabular}
\end{center}
\caption{Comparison of the number of elements $n$ of optimal povm's built using different methods as a function of the number of copies $N$ of the estimated state. The first column corresponds to the minimal solution proposed by Latorre {\em et al} in \cite{lato98} up to $N=7$. In the second column, we see that the separated variables Gauss-Legendre quadrature provides an optimal povm for arbitrary large $N$, although the number of elements $N\approx(N+1)^2/2$ is not minimal. The third column corresponds to Lebedev quadrature that reduces the number of elements to $N\approx(N+1)^2/3$ but only exists for some odd values up to $N=131$. The spherical designs, that define povm's with equally weighted elements and are known up to $N=13$, are given in the fourth column. The last two columns contains a bound on the minimal number of povm elements necessary for mixed state estimation using Gauss-Legendre quadratures and Lebedev quadratures.}\label{comparison}
\end{table}

The previous results can be straightforwardly applied to mixed states. Suppose that we are given $N$ copies of a qubit mixed state $\rho(\eta,\gras{n})=\frac{1}{2}(\gras{I}+\eta \; \gras{n} \cdot \gras{\sigma})$, where the orientation $\gras{n}$ is drawn from a uniform distribution over the $2$-sphere, and where the shrinking factor $\eta$ is drawn an arbitrary probability distribution $f(\eta)$ over the interval $[0:1]$. The problem of estimating at best the state $\rho(\eta,\gras{n})$ has been addressed  in \cite{vida99}. A key result of \cite{vida99} is a prescription of optimal state estimation strategies for mixed states from the knowledge of optimal state estimation procedures for pure states. Therefore, the povm's we have derived in the previous sections can immediately be used to obtain finite optimal povm's for mixed states. A second result of \cite{vida99}, which is relevant here is a formula for the minimum number of povm elements for optimal mixed state estimation, $n_{\textrm{min}}^{(N)}$. Interestingly, this formula is independent of the distribution $f(\eta)$. It reads:
\beq
n_{\textrm{min}}^{(N)}=\sum_{s=s_0}^{N/2} n(s),
\eeq
where $s_0=0$ if $N$ is even, and $s_0=1/2$ if $N$ is odd, $n(0)=1$ and $n(s)$ is the minimal number of povm elements for pure state estimation on $2s$ qubits.
From the pure state povm's based on the Gauss-Legendre quadrature, we find the (loose) upper bound 
$n_{\textrm{min}}^{(N)} \leq  \sum_{s=s_0}^{N/2} (2 s+1) \lceil{(2s+1)}/{2}\rceil$, from which we find
\beqa
n_{\textrm{min}}^{(N)} &\leq& \frac{N^3}{12}+\frac{5}{8}N^2+\frac{17}{12}N+1 \hspace{0.7cm} \textrm{(N even)}, \\
n_{\textrm{min}}^{(N)} &\leq& \frac{N^3}{12}+\frac{N^2}{2}+\frac{11}{12}N+\frac{1}{2} \hspace{0.7cm} \textrm{(N odd)}.
\eeqa
Furthermore, for odd values of $N$ up to $29$, we may improve on this bound by using the Lebedev quadrature,
as shown in the last column of Table~\ref{comparison}.

Finally, let us show that this approach also allows to deal with extensions of this problem,
such as the estimation of general states in the tensor product of the $N$ Hilbert spaces \cite{baga04}, instead of states
restricted to the symmetric subspace $\mathscr{H}_+^{\otimes N}$. More precisely,
suppose we want to perform state estimation on a set of states 
\bed
\{ U^{\otimes N}(g) \ket{\phi}\equiv\ket{\phi(g)}, g \in SU(2) \},
\eed
where now $\ket{\phi}$ is not restricted to belong to the symmetric subspace  $\mathscr{H}_+^{\otimes N}$.
Rather, we have 
\beq
\ket{\phi}=\sum_i a_i \ket{\phi_i},
\eeq
where the index $i$ runs over the irreducible representations of $U^{\otimes N}$.
Accordingly, $U^{\otimes N}(g)=\sum_i U_i(g)$ and
$\ket{\phi(g)}=\sum_i a_i \ket{\phi_i(g)}$, where $\ket{\phi_i(g)}\equiv U_i(g) \ket{\phi_i}$.
Suppose now that we have an optimal covariant povm for this task:
\beq
\{ U(g)^{\otimes N} \; \mathscr{O} \; U^\dagger(g)^{\otimes N} \equiv \mathscr{O}(g), g \in SU(2) \},
\eeq

The score is given by 
\bed
S=\int dg_1 \int dg_2 |\braket{g_1}{g_2}|^2 \bra{\phi(g_1)} \mathscr{O}(g_2) \ket{\phi(g_1)}
\eed
\bed
=\int dg_1 \int dg_2 |\braket{g_1 g_2^{-1}}{e}|^2 \bra{\phi(g_1 g_2^{-1})} \mathscr{O} \ket{\phi(g_1 g_2^{-1})}.
\eed

With the change of variable $g=g_1 g_2^{-1}$ and the fact that $\int dg_2=1$, we get
\bed
S=\int dg |\braket{g}{e}|^2 \bra{\phi(g)} \mathscr{O} \ket{\phi(g)}.
\eed

Again, any set of positive operators $\{c_k \mathscr{O}(g_k)\}$ that satisfies $\sum_k c_k \mathscr{O}(g_k)=\sum_i \gras{I}_i$ is an optimal povm ($\gras{I}_i$ denotes the identity over the irreducible subspace $\mathscr{H}_i$).
Since $\mathscr{H}_i=\{\ket{\phi_i(g)}, g\in SU(2) \}$,
this condition is equivalent to the following
\bed
\sum_k c_k \bra{\phi_i(g)}\mathscr{O}(g_k)\ket{\phi_i(g)}=1 \quad \forall g,i.
\eed
Let us diagonalize the povm elements as 
$\mathscr{O}=\sum_\beta \proj{\phi^{(\beta)}}$,
and define the $SU(2)$ elements $g^{(\beta)}$ such that $\ket{\phi^{(\beta)}}$ decomposes into the subspaces $\mathscr{H}_i$ as
follows
\bed
\ket{\phi^{(\beta)}}=\sum_i b_i^{(\beta)}\ket{\phi_i(g^{(\beta)})}.
\eed
We may now write the povm condition as
\bed
\sum_k c_k \sum_\beta |b_i^{(\beta)}|^2 |\bra{\phi_i(g)}U(g_k)\ket{\phi_i(g^{(\beta)})}|^2=1 \quad \forall g,i,
\eed
which will be satisfied as soon as
\beq
\sum_k c_k |\braket{\phi_i(g)}{\phi_i(g_k)}|^2=1 \quad \forall g,i.
\eeq
We therefore have a condition equivalent to
Eq.~(11) for each subspace $\mathscr{H}_i$.
Since each $\mathscr{H}_i$ corresponds to a space of total spin $S_i\leq N/2$,
we need to find a Gauss quadrature on the sphere that is exact up to order $2S_i+1$ for subspace $\mathscr{H}_i$.
It therefore suffices to consider the largest subspace, that is the symmetric subspace $\mathscr{H}_+^{\otimes N}$, corresponding to a maximal total spin $S_i=N/2$,
which requires an exact Gauss quadrature up to order $N+1$,
and taking the same quadrature for the smaller subspaces will solve the problem.
Our solution for the symmetric subspace therefore immediately yields a solution for the whole space, using a povm
with the same number of elements.

In conclusion, we have established a connection between the issue of finding optimal povm's and the problem of finding Gauss quadratures on a sphere. Following this idea, we easily derive a solution for arbitrary large $N$, that while not minimal, already improves on the best known solution \cite{derk98}. Moreover, with the help of mathematical literature on Lebedev quadrature \cite{lebe92,lebe94,lebe99} or spherical designs \cite{hard96}, we are able to give enhanced solutions for some large values up to $N=131$. Although the connection doesn't solve the problem of finding minimal optimal povm's, at least it helps to understand the difficulties involved since they turn to be essentially the same as that for finding good Gauss quadratures.
Apart from its theoretical interest, this connection could be used to design optimized measurements for experimental
state estimation. Even though the exact implementation of these povms for large number of copies $N$
would probably prove to be hard to realize experimentally,
they could be used as a starting point to derive feasible approached (non-minimal or non-optimal) povms
(see also the related works by Hayashi \cite{haya98}, and Gill and Massar \cite{gill00}).


\emph{Acknowledgements.--} We thank D. Baye, P. Capel and N. Cerf for fruitful discussions. We also thank E. Bagan for bringing to our attention the fact that a reference to Gauss-Legendre quadratures in the context of state estimation was already made in \cite{baga01}.  S.I. acknowledges financial support from the Swiss NCCR and the European project RESQ. J.R. acknowledges support from the Belgian FNRS and the French INRIA.

\emph{Note added.--} During completion of this work, we became aware of Ref. \cite{haya04} where parts of the results presented here are independently derived.

\end{document}